\documentclass[rnote,traditabstract]{aa}

\usepackage{graphicx}
\usepackage{epstopdf}
\usepackage{txfonts}
\usepackage{natbib}
\bibpunct{(}{)}{;}{a}{}{,}
\begin{document}
\title{High-precision CoRoT space photometry and fundamental parameter
  determination of the B2.5V star HD\,48977}


   \author{A.\ Thoul\inst{1,6}\fnmsep\thanks{Chercheur Qualifi\'e, Fond de
       la Recherche Scientifique (F.R.S.-FNRS)}, 
P.\ Degroote\inst{2,6}\fnmsep\thanks{Postdoctoral Fellow, Fund for
          Scientific Research of Flanders (FWO)},
C.\ Catala\inst{3}
C.\ Aerts\inst{2,4,6}, 
T.\ Morel\inst{1},
M.\ Briquet\inst{1,2}\fnmsep\thanks{Postdoctoral Fellow, Fond de
       la Recherche Scientifique (F.R.S.-FNRS)},
M.\ Hillen\inst{2},
G.\ Raskin\inst{2},
H.\ Van Winckel\inst{2},
M.\ Auvergne\inst{3},
A.\ Baglin\inst{3},
F.\ Baudin\inst{5},
E.\ Michel\inst{3}
          }

\institute{Institut d'Astrophysique et de G\'eophysique, Universit\'e de
  Li\`ege, 17 all\'ee du 6 ao\^ut, B4000, Li\`ege, Belgium\\
              \email{anne.thoul@ulg.ac.be}
\and
Instituut voor Sterrenkunde, K.U.Leuven, Celestijnenlaan 200D, B-3001 Leuven, 
Belgium
\and
LESIA, Observatoire de Paris, CNRS UMR 8109, Universit\'e Pierre et Marie Curie,
Universit\'e Denis Diderot, 5 place J. Janssen, 92105 Meudon, France
\and
Department of Astrophysics/IMAPP, Radboud University Nijmegen, 
P.O.\ Box 9010, 6500 GL
Nijmegen, The Netherlands
\and
Institut d’Astrophysique Spatiale, CNRS/Universit\'e Paris XI UMR 8617, F-091405
Orsay, France
\and
Kavli Institute for Theoretical Physics, Kohn Hall, University of California, Santa Barbara CA 93106-4030, USA
}

   \date{Received ??, 2011; accepted ??, 2011}
\authorrunning{Thoul et al.}
\titlerunning{The CoRoT light curve of the B2.5V star HD\,48977}


  \abstract
  {We present the CoRoT light curve of the bright B2.5V star HD\,48977 
    observed during a short run of the mission in 2008, as well as a
    high-resolution spectrum gathered with the HERMES
    spectrograph at the Mercator telescope. 
We use several time series analysis tools to explore the nature of the
    variations present in the light curve. 
We perform a detailed
    analysis of the spectrum of the star to determine its fundamental parameters
    and its element abundances.
We find a large number of high-order g-modes, and one rotationally induced
    frequency. We find stable 
low-amplitude frequencies in the p-mode regime as well. 
We conclude that HD\,48977 is a new Slowly Pulsating B star
 with fundamental parameters found to be $T_{\rm eff}=20\,000\pm
    1000{\rm K}$ and $\log{g}=4.2\pm0.1$. The element abundances are similar to
    those found for other B stars in the solar neighbourhood.
HD\,48977 was observed during a short run of the CoRoT satellite implying that
    the frequency precision is insufficient to perform asteroseismic
    modelling of the star. Nevertheless, we show that a longer time series of
    this star would be promising for such modelling. Our present study
    contributes to a detailed mapping of the instability strips of B
    stars in view of the dominance of g-mode pulsations in the star, several of which
    occur in the gravito-inertial regime.}


   \maketitle

%

\section{Introduction}

Asteroseismology, the study of stellar oscillations in order to infer
information about the stellar structure, is a very powerful tool to improve our
knowledge of the physics of stellar interiors \citep{book}.  Helioseismology,
the seismology of the Sun, has helped tremendously in improving our knowledge of
the Sun's structure and interior physics \citep{helioseismology}.
Asteroseismology is a younger but very dynamic field in astronomy, and several
recent space missions, in particular MOST, CoRoT, and {\it Kepler\/}, have
provided high-precision light curves for several classes of variable stars,
including main sequence B stars (e.g., \citet{aerts-most-a, aerts-most-b,
  saio2007, cameron2008, huat2009, neiner2009, diago2009, gutierrez-soto2009,
  degroote2009, degroote2010, degroote2011, papics2011, balona2011}). Despite
these high-precision data from space, in-depth seismic modelling of most of
these pulsating B stars must await (partial) identification of the wavenumbers
of the dominant observed pulsation frequencies. This was only achieved so far
for the Slowly Pulsating B 
(hereafter SPB) star HD\,50230, which led to the conclusion that this star has an
inhomogeneously mixed region surrounding its core \citep{degroote2010}, and for
the $\beta\,$Cep star HD\,180642 (V1449\,Aql) which has an unusually
high-amplitude dominant radial mode that induces several other lower-amplitude
non-radial modes of the star through resonant mode coupling \citep{briquet2009,
  hd180642}. 

In the last decade, several asteroseismic studies of B stars based on large
ground-based campaigns have been performed as well \citep{16Lac-a, 
  hd129929-a, hd129929-b, 
nueri-a, nueri-b, nueri-c, 
 nueri-e, 
nueriand12Lac, thetaophiuci-a, thetaophiuci-b,
12Lac-a, 12Lac-b}.  These observations are very extensive and
complicated, often involving multi-site multi-technique campaigns. Pulsating B
stars typically pulsate with low-order p and g modes and some also experience
high-order g modes, making it necessary to observe them for a very long time in
order to resolve all those modes. In addition, daily aliases are an
observational difficulty since these stars often have (beat) periods of the
order of a day.  Nevertheless, ground-based asteroseismic studies of B stars
have proven to be very successful and have led to important new insights in
their interior physics.  For example, it was possible to put limits on the value
of the overshooting parameter of the core \citep{nueri-d, hd129929-c, thetaophiuci-c,
  12Lac-b}.  From partially resolved multiplets due to rotational
splitting, it was also possible to show that some of those stars have non-rigid
rotation, in the sense that their core regions rotate faster than their envelope
\citep{pamyatnykh2004,nueri-d,nueriand12Lac}.  Finally, the seismic modelling of those B stars
provided values for their fundamental parameters, such as the effective
temperature, $\log\,g$ and age.

In this paper we are concerned with data
assembled with the CoRoT (\citet{corot}) satellite. 
Its main asteroseismology
mission is to observe a limited number of bright main sequence stars during a
long time span of maximum five months, but in between those long runs also
shorter ones of some three to five weeks occur.  Because of the success
of the ground-based asteroseismology of B stars, several B stars with a variety
of spectral type were selected as targets for CoRoT, with the aim to map the
instability strips of the $\beta\,$Cep and SPB stars.

This paper is devoted to HD\,48977, a B2.5V star ($V$ magnitude of 5.92) which
was observed by CoRoT as a secondary asteroseismology target during a short run
in 2008.  Its spectral type of B2.5V places it in the joint region of the SPB
and $\beta\,$Cep instability strip.  Very little was known about this star prior
to the CoRoT observations. It had been flagged as a potential rotationally or
pulsationally variable star and it was observed by Hipparcos, which led to a
main frequency peak at 0.5218\,d$^{-1}$ \citep{hipparcos2002}.  Estimates of its
fundamental parameters using colour indices were given in \citet{lyubimkov2002}
as $T_{\rm eff}=17500\pm 300$, $\log\,g=4.09\pm 0.08$, from which the authors
deduced $M=5.7\pm 0.3\,M_{\odot}$, $R=3.6\pm 0.4R_{\odot}$, an age of $30.6\pm
10.7 \times 10^6$\,yrs, and d=$252\pm 40 {\rm pc}$ using \citet{claret1995}
evolutionary tracks.


\section{CoRoT high-precision photometry of HD\,48977}

\begin{figure}
\centering
\includegraphics[width=\columnwidth]{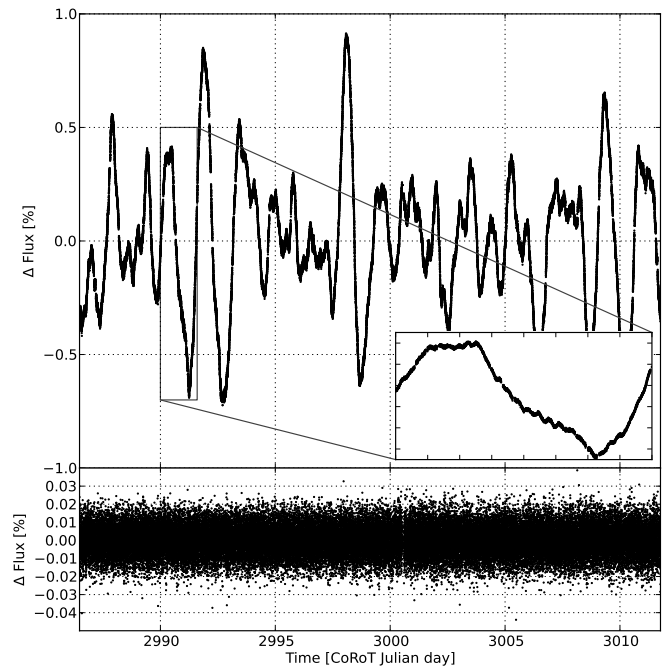}
\caption{CoRoT light curve of HD\,49877, with an inset to show the
  short-time behaviour. The bottom panel represents the residuals after
  prewhitening with a model based on the 55 frequencies listed in
  Table\,\protect\ref{frequencies}.} 
\label{lc}
\end{figure}

\begin{figure}
\centering
\includegraphics[width=\columnwidth]{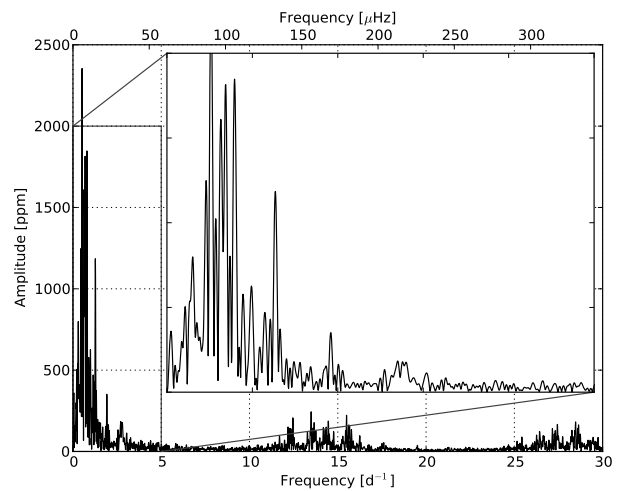}
\caption{Fourier transform of the light curve.} 
\label{FT}
\end{figure}


HD\,48977 was observed by CoRoT with a cadence of 32\,s during a short run in
2008, from 2454531.48027\,JD (5 May) to 2454556.76361 (31 May), i.e., during
some 25 days.  The detrended light curve is shown in Fig.~\ref{lc}.  It was
obtained by fitting the light curve with a linear polynomial and dividing the
light curve by this polynomial in order to convert to flux units and compensate
for the loss of gain in the satellite CCD \citep{auvergne2009}.  

Searching for
frequencies in the light curve was done independently by AT using Period04
\citep{period04} and by PD using the method in Degroote et al.\ (2009).  
The Fourier transform is shown in Fig.~\ref{FT}. It can be seen that the dominant signal 
occurs in the range below 3\,d$^{-1}$, which is typical for g-mode pulsations in early B stars.
The two
methods give the same list of frequencies (within the resolution).  The
resulting 55 frequencies found having an amplitude more than 4 times the noise
level are listed in Table~\ref{frequencies}.  The main peak has a frequency of
0.5167\,d$^{-1}$, very close to the frequency 0.5218\,d$^{-1}$ of the Hipparcos
light curve \citep{hipparcos2002}.  We see that HD\,48977 presents a rich
spectrum dominated by g modes with frequencies
between 0.1\,d$^{-1}$ and 1.5\,d$^{-1}$. 
We thus classify HD\,48977 as a new SPB
star. 

An amplitude drop with a factor more than two occurs for the frequencies 35 to
55 in Table~\ref{frequencies}, which represent power in the p-mode regime,
compared to the lower frequency regime (Fig.~\ref{FT}).  While these 20 listed
frequencies above 10\,d$^{-1}$ are significant, we are cautious with their
interpretation because they occur in the region where instrumental power also
occurs (\citet{auvergne2009}, see also Fig.\ref{FT}).  In particular, the two
highest frequencies in Table~\ref{frequencies} are twice and five times the
orbital frequency of the satellite. Four additional peaks in
Table~\ref{frequencies} coincide with (daily aliases of) the satellite orbital
frequency, keeping in mind the relatively large frequency uncertainty.
Nevertheless, significant intrinsic stellar frequencies occur in the region
[10,20]\,d$^{-1}$. As is also the case for the pulsators HD\,50230 (B3V,
\citet{degroote2010}) and HD\,43317 (B3IV, \cite{papics2012}), the
frequency structure in that regime is rather stable 
pointing
towards p modes excited by the kappa mechanism.  The autocorrelation of the
power spectrum 
peaks at a possible frequency separation of 1.55\,d$^{-1}$, but, in
view of the limited frequency precision and the unknown identification of the
spherical wavenumbers of the modes, we refrain from further interpretation of
the frequencies in this regime with the present data set at hand.  The noise
level in the residual amplitude spectrum, measured between 5 and 10\,d$^{-1}$,
amounts to only 1 ppm.

\begin{figure}
\centering
\includegraphics[width=\columnwidth]{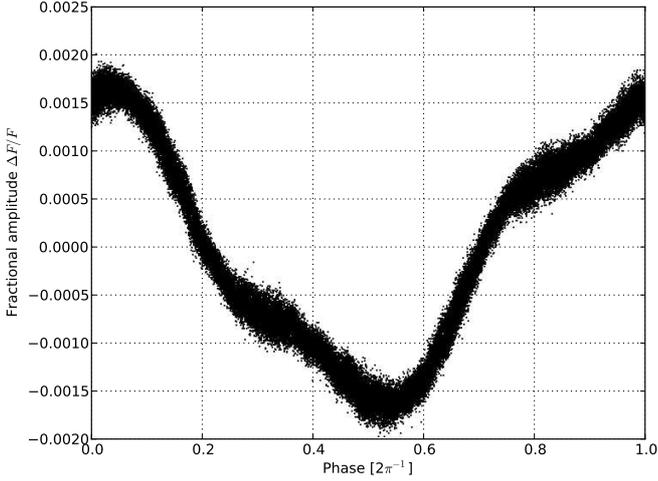}
\caption{Phase plot of the rotational signal, where all signal not related to the rotational frequency and its harmonics has been filtered out.}
\label{phase}
\end{figure}

Given that the harmonics of the fourth frequency $f_4=0.637\,$d$^{-1}$ are
clearly present in the frequency spectrum, while there are no harmonics for any
of the other frequencies, and given that the shape of the curve in the phase diagram is not typical of nonlinear 
pulsations (see Fig.~\ref{phase})
we interpret $f_4$ as the rotation frequency of the
star. Also this particular aspect of the frequency spectrum is similar for the
B3IV pulsator HD\,43317 (\cite{papics2012}), although the data set
of the latter star is much more conclusive on this matter, given that this
target was observed during a long run of CoRoT.  

\begin{table}
  \caption{Frequencies in the light curve of HD\,48977. SN2 is the signal-to-noise in a 6\,$d^{-1}$ bin around the peak after each prewhitening step. Phases are
given with respect to $t_0=0$, and in units of $2\pi\rm{rad}$.}
\label{frequencies}
\centering
\begin{tabular}{ cccccccc }
\hline\hline
 nr & $A$ & $\sigma(A)$& $f$ & $\sigma(f)$     & $\phi$ &  $\sigma(\phi)$ &  SN2\\
    & (ppm)           & (ppm)      & (d$^{-1}$)       & (d$^{-1}$)      & (2$\pi$)          & (2$\pi$)     &     \\\hline
  1 & 2052        &   125 		&   0.5167       &      0.0011 &       0.38 &      0.05 &   14.7\\
  2 & 1737         &   109 		&   0.7933       &      0.0013 &    -0.085 &       0.06 &   12.6\\
  3 & 1583         &    89 		&   0.6832       &      0.0012 &    -0.019 &      0.06 &   12.8\\
  4 & 1489         &    66 		&   0.6370       &      0.0010 &      0.26 &      0.04 &   13.5\\
  5 & 1130         &    44 		&   1.2688       &      0.0008 &      0.24 &       0.04 &  12.0\\
  6 & 924         &    34 		&   0.4542       &      0.0009&      0.37 &      0.04 &   9.3 \\
  7 & 753        &    38 		&   0.3049       &     0.0012 &     0.07 &      0.05 &    9.1 \\
  8 & 647         &    27 		&   0.9847       &     0.0009 &     0.06 &      0.04 &    9.0 \\
  9 & 498        &    19 		&   0.8958       &     0.0010 &     -0.46 &      0.04 &   7.1 \\
 10 & 476         &    21 		&   1.1412       &     0.0010 &     -0.42 &      0.04 &   7.4\\
 11 & 430         &    17.		&   0.5699      &     0.0010 &      0.29 &       0.04 &   7.1 \\
 12 & 427        &    26 		&   0.2100       &       0.0011 &      0.40 &        0.05 &   7.5\\
 13 & 393         &    14 		&   0.3950       &     0.0009 &      0.19 &      0.04 &   6.7 \\
 14 & 388         &    17 		&   0.2604       &     0.0010 &     -0.28 &      0.04 &   7.2 \\
 15 & 330         &    12 		&   0.3484       &     0.0009 &     -0.24 &      0.04 &   6.8 \\
 16 & 299         &    11 		&   1.9182       &     0.0008 &     -0.31 &      0.04 &   6.7 \\
 17 & 292         &     9 		&   0.0410       &     0.0007 &      0.15 &       0.03 &   6.6 \\
 18 & 201         &     8 		&   1.0766      &      0.0011 &     -0.49&      0.05 &   4.5 \\
 19 & 188         &     9		 &   2.5325       &      0.0010 &     -0.33 &      0.05 &   4.7 \\
 20 & 182        &     8		 &   0.1729       &      0.0010 &      0.36 &      0.05 &   4.3 \\
 21 & 175        &     7 		&    2.7611      &      0.0010&     -0.49 &      0.05 &   4.2 \\
 22 & 152         &     6 		&   0.6008       &      0.0010 &     -0.33 &      0.05 &   4.2 \\
 23 & 143         &     7 		&    2.7110       &      0.0010 &     -0.12 &      0.05 &   4.1 \\
 24 & 142         &     6 		&   0.8440       &     0.0010 &     -0.31 &       0.04 &   4.3 \\
 25 & 127          &     6 		&    1.3232      &      0.0010 &     -0.42 &      0.05 &   4.0 \\
 26 & 125         &     7 		&    2.0672       &       0.0010 &     -0.48 &      0.05 &   4.1 \\
 27 & 124         &     6 		&    1.7022       &      0.0010 &     -0.36 &      0.05 &   4.1 \\
 28 & 123         &     5 		&    1.8443       &      0.0010 &     -0.33 &      0.05 &   4.1 \\
 29 & 115         &     5 		&    3.0379       &     0.0010 &      0.36 &      0.04 &   4.1 \\
 30 & 112         &     5 		&    1.0288       &      0.0010 &     -0.49 &      0.05 &   4.0\\
 31 & 108         &     5 		&   0.7579       &     0.0010 &     -0.30 &      0.04 &   4.1 \\
 32 & 107         &     4 		&    2.6192       &     0.0010 &     -0.45 &      0.04 &   4.0 \\
 33 & 104         &     4		&    2.9035      &     0.0010 &     -0.33 &       0.04 &   4.0 \\
 34 & 94         &     4 		&    2.8083       &     0.0009 &     0.05 &      0.04 &   4.2 \\
 35 & 44         &     2.5		 &    14.7599       &      0.0013 &     -0.20 &      0.06 &   7.1 \\
 36 & 43         &     2.5 		&    13.4966       &      0.0013 &      0.42 &      0.06 &   6.4\\
 37 & 43         &     2.5 		&    14.4568       &       0.0013 &     -0.23 &      0.06 &   7.9 \\
 38 & 33         &     1.9 		&    16.3164       &      0.0013 &     -0.36 &      0.06 &   6.7 \\
 39 & 29         &     1.8 		&    11.1543       &      0.0013 &     -0.49 &      0.06 &   6.4 \\
 40 & 22         &     1.5 		&    19.4616       &      0.0014 &     -0.48 &      0.06 &   5.6 \\
 41 & 22         &     1.5 		&     10.484       &      0.0014 &     -0.41 &      0.06 &   4.5 \\
 42 & 21         &     1.4 		&    17.9453       &      0.0014 &     -0.25 &      0.06 &   4.8 \\
 43 & 21         &     1.4 		&    15.7987       &      0.0014 &     0.069 &      0.06 &   5.0 \\
 44 & 20        &     1.3 		&    14.6173       &      0.0014 &      0.20 &      0.06 &   5.5 \\
 45 & 19         &     1.3 		&    19.5002       &      0.0014 &     -0.18 &      0.06 &   5.8 \\
 46 & 18         &     1.1 		&    15.6412       &      0.0014 &     0.097 &      0.06 &   4.7 \\
 47 & 18         &     1.2 		&    17.6694       &      0.0014 &     -0.14 &      0.06 &   4.9 \\
 48 & 17        &     1.2 		&    16.0906       &      0.0014&     0.0027 &      0.06 &   4.7 \\
 49 & 17         &     1.0 		&     12.872       &      0.0013 &      0.48 &      0.06 &   4.8 \\
 50 & 17        &     1.1 		&    11.0296       &      0.0014 &      0.15 &      0.06 &   4.1 \\
 51 & 14         &     0.9 		&    21.2714       &      0.0014 &      0.11 &      0.07 &   5.3\\
 52 & 14        &     0.9 		&    15.9682       &      0.0014 &      0.32 &      0.06 &   4.3 \\
 53 & 9.5         &     0.7 		&    22.9226       &       0.0017 &      0.13 &      0.08 &   4.7 \\
 54 & 8.5         &     0.7 		&    27.9437       &      0.0017 &      0.47 &      0.08 &   7.1 \\
 55 & 3.2        &     0.5 		&    69.8564       &      0.0034 &      0.33 &       0.15 &   5.3 \\
\hline
\end{tabular}
\end{table}

The frequency resolution deduced from the CoRoT light curve is given by
$1.5/\Delta T\simeq 0.06\,$d$^{-1}$, where $T$ is the entire duration of the
observation run, i.e., 25 days.  Unfortunately, this is of the order of the
separation between the observed g-mode 
frequencies (see Table\,\ref{frequencies}).
This prevented us from detecting any meaningful period spacings, as deduced for
HD\,50230 by Degroote et al.\ (2010), due to the fact that those spacings would
be of the same order as the frequency resolution itself.  

The frequency spectrum we have at hand is of too low precision to perform a
meaningful seismic modelling of this star. This is not surprising, as SPB stars
are known to need observations of several months before the frequency spectrum
can be deduced with sufficient precision to interpret it in terms of predicted
frequencies from theoretical models. The selection of this secondary seismology
target for a short run was based on its rough position in the instability
strips, which could have implied the detection of dominant 
high-amplitude p modes. A good strategy for
further understanding the target would be to re-observe it with CoRoT during a
medium to long run of several months.


\section{Spectroscopy of HD\,48977}

\begin{figure}
\centering
\includegraphics[width=\columnwidth]{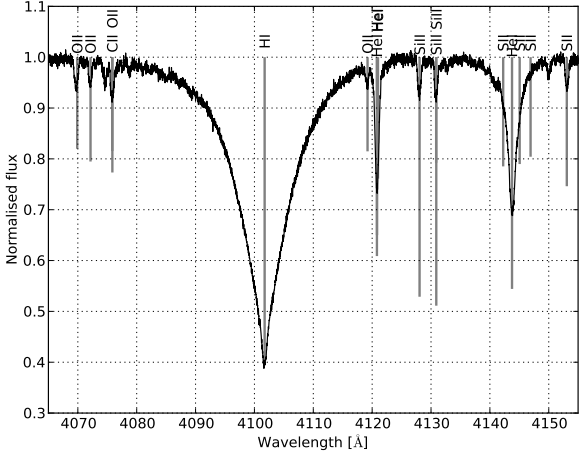}
\caption{Excerpt of the high resolution high signal-to-noise spectrum (black) of
  HD\,48977, obtained with the HERMES spectrograph attached to the Mercator
  Telescope. Grey lines indicate the zero-rotation line depths of a template
  spectrum with $T_\mathrm{eff}=20\,000\,K$ and $\log{g}=4.2$\,dex (solar
  mixture and solar metallicity).  }
\label{hermes}
\end{figure}

One high resolution high signal-to-noise spectrum of HD\,48977 was obtained
using the HERMES spectrograph \citep{hermes2011} at the Mercator Telescope.  The normalised
spectrum for the spectral range 4065-4155\AA\ is shown in Fig.~\ref{hermes}.  It
has a resolving power of 80\,000 and a S/N ratio of about 170 in the region near
the Si\,II doublet at 4128, 4130\AA\ which is a spectral range most often used
to study the oscillations of SPB stars \citep{aertsetal99}. From the Fourier
transform of the SiIII\,4567 line, which allows the disentangling of
non-rotational broadening from other broadening mechanisms \citep[i.e.,
pulsations, micro- and macroturbulence, ][]{simondiaz2007}, we deduce $v\sin\,i
= 29\pm 1\,$km\,s$^{-1}$. Combined with the interpretation of $f_4$ as the
rotation frequency and a canonical radius of B-type stars in this part of the Hertzsprung-Russell
diagram ($R=3.5\pm0.7\,R_\odot$, obtained from a grid of models calculated using
CLES \citep{cles}), we derive an inclination angle $i=15\pm3^\circ$, or a true
equatorial velocity $v_\mathrm{eq}=113\pm23$\,km\,s$^{-1}$.  


We performed a detailed analysis of the spectrum of HD\,48977 using the method
explained in \citet{morel2006} to determine the fundamental parameters and the
element abundances of the star. The synthetic spectra used have been computed
using an updated version of the non-LTE line-formation codes DETAIL/SURFACE
(\cite{butler_giddings}; \cite{giddings})

\begin{table}
  \caption{Si lines
 used to determine the effective temperature in B stars. 
 Lines present in the spectrum of HD\,48977 are underlined. 
Lines that should be avoided according to \citet{simondiaz2010} 
are listed in italic.}            
\label{Si-lines}      
\centering                       
\begin{tabular}{c c c}        
\hline\hline                 
Si II  & Si III  &Si IV  \\    
\hline                       
   \underline{\it 4128.054} & \underline{4567.840} & 4212.405\\      
   \underline{\it 4130.894} & \underline{4574.757} & \it{6701.12} \\
   \underline{\it 5056.150} & \it{4813.33}               &\\
   \underline{6371.371}     & \it{4829.073}             &\\
                                       & 5716.287                    &\\
                                       & \underline{5739.734}  &\\
\hline                                  
\end{tabular}
\end{table}

\begin{figure}
\centering
\includegraphics[width=\columnwidth]{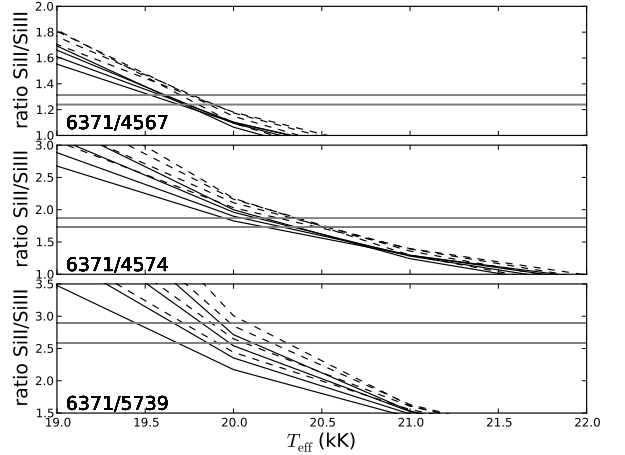}
\caption{Theoretical EW ratios for $\mathrm{log}\,g=4.2$ of the suitable Si\,II
  and Si\,III lines in the spectrum with proper atomic data.  Theoretical
  values are plotted for two values of the microturbulence,
  $\xi=1\,$km\,s$^{-1}$ (full lines) and $\xi=5\,$km\,s$^{-1}$ (dashed lines),
  and four values of the Si abundance, $\mathrm{log} \epsilon$(Si)=7.54, 7.34,
  7.14, 6.94\,\rm{dex}. The two horizontal lines represent the observed values
  with their error bar.}
\label{Si-ratios}
\end{figure}

After normalizing the spectrum, we measured 
the ratios of the equivalent widths (EWs) of the
different ionization lines of Si, in order to obtain rough estimates of the
effective temperature. Some Si lines appropriate for use in the case of B stars
are listed in Table~\ref{Si-lines}.  Among those lines, all four Si\,II lines
are present in the spectrum of HD\,48977, while only three Si\,III lines are
present and no Si\,IV lines. The lines present in the spectrum of HD\,48977 are
underlined in Table~\ref{Si-lines}. Following \citet{simondiaz2010}, some of
those lines are badly understood from the viewpoint of atomic data and are
better avoided in the analysis. Those lines are indicated in italic in
Table~\ref{Si-lines}.  Looking at Table~\ref{Si-lines}, we see that we can use
only one Si\,II line and three Si\,III lines. The ratios of the EWs of the
Si\,II and Si\,III lines are shown in Fig.~\ref{Si-ratios} for $\log\,g=4.2$.
The theoretical ratios are shown for two values of the microturbulence, namely
$\xi=1\,$km\,s$^{-1}$ and $\xi=5\,$km\,s$^{-1}$ and for four values of the Si
abundance, $\log \epsilon$(Si)=7.54, 7.34, 7.14, 6.94\,\rm{dex}.  This gives an
estimate for the effective temperature of $T_{\rm eff}=19\,900\pm 400\,{\rm K}$.
Given the fact that we have so few suitable Si lines, 
and for completeness, we did
the same analysis using all the observed Si lines. From the lines that should
not be used according to \citet{simondiaz2010}, we get higher values for the
effective temperature, up to 22\,000 K.  This is consistent with the general
trend reported by \citet{simondiaz2010}.

\begin{figure}
\centering
\includegraphics[width=\columnwidth]{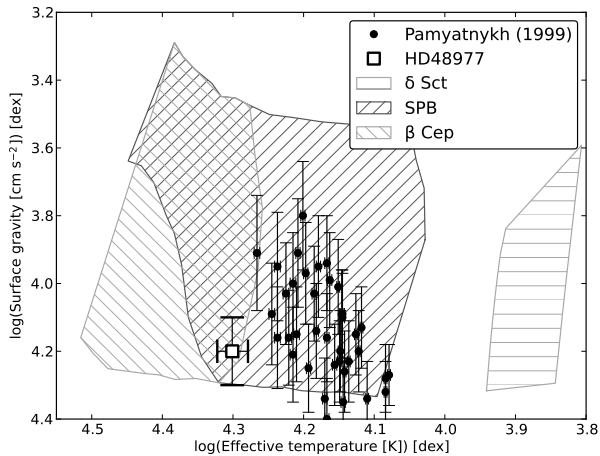}
\caption{Instability strips for B stars, as well as observed SPB stars from
  ground-based surveys (dots) taken from \citet{Pamyatnykh99}. The position of
  HD\,48977 is indicated as square.}
\label{instab}
\end{figure}

The next step is to find the best value for $\log\,g$ by fitting the wings of
the Balmer lines. We use the H$_\beta$, H$_\gamma$, H$_\delta$, H$_\epsilon$,
and H$_\alpha$ lines. We find the best fit for $\log\,g=4.2\pm 0.1$ assuming
$\log\,T_{\rm eff}=20 000\pm 1000\,$K.  
In Fig.~\ref{instab}, we show
HD\,48977 in the B star instability strips. We see that it falls nicely in the
SPB instability region.

In principle, the value for the microturbulence parameter can be found by
looking at strong and weak lines of a given ion, 
and require that they yield the same abundances.
Unfortunately, all O lines in the spectrum of HD\,48977 are weak and
we cannot determine the value of the microturbulence. It is, however, expected
to be small for this spectral type.  We included the unknown value of the
microturbulence parameter for the element abundances by increasing the error
bars based on two possible values of $\xi=1\,$km\,s$^{-1}$ and
$\xi=5\,$km\,s$^{-1}$.

To refine the estimate of the effective temperature, we required the Si\,II and
Si\,III lines to return on average the same abundance.  Using the Si lines
underlined in Table\,\ref{Si-lines}, we obtain a value consistent with the
previous estimate and $\log \epsilon$(Si)=7.27$\pm$0.20\,dex.

We can now obtain the abundances of the other elements with lines in the
spectrum. The lines used for the analysis and the results obtained for the
abundances are summarized in Table~\ref{abundances}.  The abundance
uncertainties take into account both the line-to-line scatter and the errors
arising from the uncertainties in the atmospheric parameters.  They are in very
good agreement with the abundances found for O and B stars in the solar
neighourhoud by \citet{morel2006} using exactly the same methods and tools.  The
Ne abundance is found to be 7.96 dex, a value intermediate between the solar
values recommended by \citet{GS98} and \citet{AGS09}.  
\citet{morel2008} obtained a very
similar Ne abundance for B stars.

The CNO abundance ratios are found to be [N/O]=-0.92 and [N/C]=-0.61, which are
equal to the solar values within the errors.  We thus do not find any indication
of deep mixing in HD\,48977. Such mixing was found for several among the bright
$\beta\,$Cep stars \citep{morel2006}, which are somewhat hotter than HD\,48977.

\begin{table*}               
  \caption{Non-LTE elemental abundances of \object{HD 48977} on the scale in which $\log \epsilon$[H]=12. 
    Mean values obtained using the same techniques for nearby, early B-type stars showing no signs of mixing (
    i.e., with unaltered CNO abundances) are shown for comparison \citep[][]{morel2008, moreletal2008}.}            
\label{abundances}      
\centering                       
\begin{tabular}{l l l c c}        
\hline\hline                 
element  & ion used & lines used (\AA)    & \object{HD 48977}  & Nearby B stars\\    
\hline                       
C    &  C\,II   &5133.11                                                      &8.14$\pm$0.10        &8.21$\pm$0.09\\
N    &  N\,II   &3994.00, 4607.16, 4643.09, 5005.15, 5679.55                  &7.53$\pm$0.12        &7.67$\pm$0.11\\
O    &	O\,II   &4069.75, 4414.90, 4641.81, 4649.13, 4661.63, 4705.35         &8.45$\pm$0.12        &8.51$\pm$0.09\\
Ne   &  Ne\,I   &6143.06, 6266.49, 6334.43, 6382.99, 6402.25, 6506.53         &7.96$\pm$0.12        &7.97$\pm$0.07\\
Mg   &  Mg\,II  &7877.05                                                      &7.61$\pm$0.10        &7.39$\pm$0.11\\
Al   &  Al\,III &4149.96, 4512.54, 4529.04, 5696.6, 5722.7                    &6.21$\pm$0.28        &6.13$\pm$0.08\\
Si   &  &see Table~\ref{Si-lines}                                             &7.27$\pm$0.20        &7.18$\pm$0.07\\
S    &  S\,II   &4162.48, 4524.81, 4815.55                                    &7.11$\pm$0.18        &7.20$\pm$0.11\\
Fe   &  Fe\,III &4419.60, 5156.11                                             &7.21$\pm$0.14        &7.28$\pm$0.10\\
\hline                                  
\end{tabular}
\end{table*}


\section{Photometric determination of the effective temperature}

The visual magnitude of HD\,48977 makes it an optimal target for inclusion in
wide-field surveys. We collected absolute photometry in the UV \citep[TD1, ANS,
][]{cat_td1,calib_ans,cat_ans}, the optical \citep[Str{\o}mgren, Geneva, Johnson
and Tycho][]{cat_gcpd,cat_geneva,calib_geneva,cat_pasp,cat_ubv,cat_tycho} and
the (near) infrared \citep[Johnson, 2MASS, WISE, AKARI,
][]{cat_2mass,cat_wise,calib_wise,cat_akari}. From the WISE bands at $3.4\,\mu$m
(W1), $4.6\,\mu$m (W2), $12.3\,\mu$m (W3) and $22\,\mu$m (W4) and AKARI
photometry at 8.7\,$\mu$m, we see no signs of infrared excess.

\begin{figure}
 \includegraphics[width=\columnwidth]{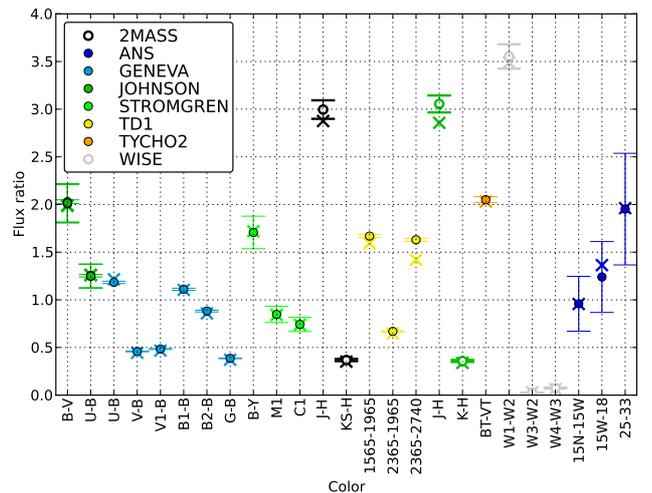}
 \caption{Observed and model photometric colors of HD\,48977. Crosses denote the
   synthetic flux of the best fitting model, circles denote the observations and
   their errors. Filled circles are included in the fit as colors (optical and
   UV), open circles are included as absolute photometry (infrared).}
\label{fig:colors}
\end{figure}

Even though it is well known that fundamental parameters determined from SEDs based on 
photometric data are usually less precise than those deduced from high-resolution spectroscopy,
we estimated the effective temperature and gravity via the procedure outlined in \citet{degroote2011}, 
as a compatibility check. We
chose to fit the absolute fluxes in the infrared ($\lambda>1\,\mu$m), and
only colours at shorter wavelengths \citep[cf. the infrared flux method][]{irfm}
(Fig.\,\ref{fig:colors}). We used both LTE \citep{castelli2003} and NLTE model
atmospheres \citep{lanz2007}. We deduced an effective temperature
$T_\mathrm{eff}=18\,000\pm1\,300$\,K (95\% confidence interval), surface gravity
$\log\,g>3.6$, interstellar exctinction $E(B-V)<0.025$, angular diameter
$\theta=0.128\pm0.004$\,mas and a solar or sub-solar metallicity. We thus find
that the photometric method underestimates the effective temperature by roughly
$1\,000$\,K compared to the silicon line ratios (Fig.\,\ref{Si-ratios}) but the
values are consistent within 2$\sigma$. Given the much higher quality of the HERMES spectrum
compared to broad-band photometry, we adopt the spectroscopic estimate of the effective temperature.

\section{Conclusions}

From 25\,d of CoRoT high-precision space photometry, we found the B2.5V star
HD\,48977 to be a new SPB star. In fact, it is the hottest confirmed SPB star
known to date.  In agreement with the star's position in the joint part of the
$\beta\,$Cep and SPB instability strips, we detected numerous dominant g modes
and low-amplitude p modes in its oscillation spectrum.  The g modes occur in the
frequency range 0.17 -- 3.0 \,d$^{-1}$ and reach a maximum amplitude of
2000\,ppm.  The ratio of the p- and g-mode amplitudes of this moderately
rotating star of solar metallicity is less than 0.025. In this aspect, the three stars HD\,48977 (B2.5V, moderate rotator;
this paper), HD\,50230 (B3V, slow rotator; \citet{degroote2010}) and HD\,43317
(B3IV, fast rotator; \cite{papics2012}) are very comparable.

While the frequency precision is insufficient to perform forward modelling at
this stage, an interesting aspect of HD\,48977's variability is that we detected
rotational modulation, in addition to the g-mode oscillations, in its
photometric light curve, again similar to the case of HD\,43317.  The
combination of the rotation frequency we found in the CoRoT light curve and the
overall line broadening in the spectrum, implies that HD\,48977 is one of the
very few moderate rotators known within the class of SPB stars
so far, the B3V star HD\,43317 being an even faster rotator (\cite{papics2012}).

The rotation frequency of 0.6372\,d$^{-1}$ is of the same order as the g-mode
frequencies and implies that any future seismic modelling of the star must take
into account the effects of rotation in the pulsation description.  While this
can be done using the traditional approximation for modes whose frequency is
above twice the rotation frequency \citep{aertsdupret11,ballot11}, the
lowest-frequency modes of HD\,48977 correspond with so-called gravito-inertial
waves and require a treatment not relying on perturbative methods
\citep{ballot10}.

Our findings show that HD\,48977 would be an excellent target for a long run
during an extension of the CoRoT mission. This would probably allow to decide on
the nature of the power in the frequency regime above 10\,d$^{-1}$ and to
unravel rotational splitting of several of the pulsation modes, and hence would
open stellar modelling applications from the mode identification of the
multiplet structures.


\begin{acknowledgements}
  A.T., P.D., and C.A., are grateful to the staff of the Kavli Institute of
  Theoretical Physics (KITP) of the University of California at Santa Barbara
  (UCSB) for the hospitality during their stay in the framework of the 2011
  Research Programme ``Asteroseismology in the Space Age''. 
T.M. acknowledges financial support from Belspo for contract PRODEX
  GAIA-DPAC. The research leading to these results
  has received funding from the European Research Council under the European
  Community’s Seventh Framework Programme (FP7/2007–2013)/ERC grant agreement
  n$^{\rm o}$227224 (PROSPERITY), from the Belgian Science Policy Office
  (Belspo) under PRODEX contract C90309: CoRoT Data Exploitation, and from the
  National Science Foundation of the United States under grant n$^{\rm o}$NSF
  PHY05--51164, and from the FNRS.
The CoRoT space mission was
    developed and is operated by the French space agency CNES, with
    participation of ESA’s RSSD and Science Programmes, Austria, Belgium,
    Brazil, Germany, and Spain.  Based on observations made with the HERMES
    spectrograph, installed at the Mercator Telescope, operated on the island of
    La Palma by the Flemish Community, at the Spanish Observatorio del Roque de
    los Muchachos of the Instituto de Astrof\'{\i}sica de Canarias and supported
    by the Fund for Scientific Research of Flanders (FWO), Belgium, the Research
    Council of K.U.Leuven, Belgium, the Fonds National de la Recherche
    Scientific (F.R.S.--FNRS), Belgium, the Royal Observatory of Belgium, the
    Observatoire de Gen\`eve, Switzerland and the Th\"uringer Landessternwarte
    Tautenburg, Germany. 
\end{acknowledgements}

\bibliographystyle{aa}

\bibliography{hd48977-RN}

\end{document}